\documentclass[aps,prl,twocolumn,superscriptaddress,showpacs]{revtex4-1}
\usepackage{graphicx}
\usepackage{amsmath} 
\usepackage{hyperref} 
\usepackage{color}
\usepackage[utf8]{inputenc}

\begin{document}


\title{Long-term monitoring of the internal energy distribution of isolated cluster systems}
\author{Christian Breitenfeldt}
\affiliation{Institut f\"ur Physik, Ernst-Moritz-Arndt-Universit\"at, 17487 Greifswald, Germany}
\affiliation{Max-Planck-Institut f\"ur Kernphysik, 69117 Heidelberg, Germany}
\author{Klaus Blaum}
\affiliation{Max-Planck-Institut f\"ur Kernphysik, 69117 Heidelberg, Germany}
\author{Sebastian George}
\affiliation{Max-Planck-Institut f\"ur Kernphysik, 69117 Heidelberg, Germany}
\author{J\"urgen G\"ock}
\affiliation{Max-Planck-Institut f\"ur Kernphysik, 69117 Heidelberg, Germany}
\author{Gregorio Guzm\'an-Ram\'irez}
\affiliation{Departamento de Ingenier\'ias, Centro Universitario de Tonal\'a, Universidad de Guadalajara, Jal.\ 48525, Mexico}
\author{Jonas Karthein}
\affiliation{Max-Planck-Institut f\"ur Kernphysik, 69117 Heidelberg, Germany}
\author{Thomas Kolling}
\affiliation{Fachbereich Chemie and Forschungszentrum OPTIMAS, Technische Universit\"at Kaiserslautern, 67663 Kaiserslautern, Germany}
\author{Michael Lange}
\affiliation{Max-Planck-Institut f\"ur Kernphysik, 69117 Heidelberg, Germany}
\author{Sebastian Menk}
\affiliation{Max-Planck-Institut f\"ur Kernphysik, 69117 Heidelberg, Germany}
\author{Christian Meyer}
\affiliation{Max-Planck-Institut f\"ur Kernphysik, 69117 Heidelberg, Germany}
\author{Jennifer Mohrbach}
\affiliation{Fachbereich Chemie and Forschungszentrum OPTIMAS, Technische Universit\"at Kaiserslautern, 67663 Kaiserslautern, Germany}
\author{Gereon Niedner-Schatteburg}
\affiliation{Fachbereich Chemie and Forschungszentrum OPTIMAS, Technische Universit\"at Kaiserslautern, 67663 Kaiserslautern, Germany}
\author{Dirk Schwalm}
\thanks{deceased 14 July 2016}
\affiliation{Max-Planck-Institut f\"ur Kernphysik, 69117 Heidelberg, Germany}
\affiliation{Department of Particle Physics, Weizmann Institute of Science, Rehovot 76100, Israel}
\author{Lutz Schweikhard}
\affiliation{Institut f\"ur Physik, Ernst-Moritz-Arndt-Universit\"at, 17487 Greifswald, Germany}
\author{Andreas Wolf}
\affiliation{Max-Planck-Institut f\"ur Kernphysik, 69117 Heidelberg, Germany}

\begin{abstract}
  A method is presented to monitor the internal energy distribution of cluster anions via delayed electron detachment by pulsed
  photoexcitation and demonstrated on Co$_4^-$ in an electrostatic ion beam trap.  In cryogenic operation, we calibrate the detachment
  delay to internal energy.  By laser frequency scans, at room temperature, we reconstruct the time-dependent internal energy
  distribution of the clusters.  The mean energies of ensembles from a cold and a hot ion source both approach thermal equilibrium.
  Our data yield a radiative emission law and the absorptivity of the cluster for thermal radiation.
\end{abstract}

\maketitle

Understanding of complex molecules and clusters is governed by the interplay of microscopic description and multistate statistics.
Statistical methods were successfully applied to unimolecular reactions (such as dissociation or electron emission)
\cite{andersen_thermionic_2002} and radiative interactions \cite{hansen_thermal_1998}.  Near infrared radiation from nanosystems is
important for understanding interstellar continuum emission \cite{duley_excitation_2009} as well as the survival of molecular matter in
the interstellar medium, such as carbon bearing anions \cite{herbst_calculations_2008,joblin_polycyclic_1997}.  Spectral observations
in some wavelength regions point to continuum black-body emission even from very small clusters \cite{mitzner_optical_1995}. On the
other hand, it is still not well understood how the emission properties are related to the particles' internal energy distributions
\cite{fan_probing_2009}, even at the higher temperatures investigated so far \cite{mitzner_optical_1995,fan_probing_2009}.

Recently, ion-trap experiments could access internal energy relaxation through studies of delayed electron emission from molecular and
cluster anions.  Time-resolved measurements indirectly demonstrated Stefan--Boltzmann-like relaxation \cite{toker_radiative_2007} and
sampled shifts of internal energy distributions for up to $\sim$100\,ms \cite{goto_direct_2013,najafian_radiative_2014}.  Yet, basic
uncertainties remain: Without direct scanning of the internal energy distribution (IED) in a cluster ensemble it is unclear how
closely the IEDs follow a canonical shape.  Moreover, radiative relaxation was studied only for cluster temperatures high above 300\,K
\cite{walther_radiative_1999,hansen_thermal_2014}.  Much less is known at longer observation times and for temperatures where
far-infrared emission dominates.

In the present work, we succeed in tracking IEDs in small cluster anions stored in vacuum over several seconds, until the distributions
from a hot and a cold ion source both approach thermal equilibrium with the environment.  This is achieved by a direct bin-wise
measurement of the IED using laser excitation.  Single-photon excitation of stored anions by a nanosecond laser pulse increases their
internal energy $E$ by the photon energy $h\nu$ from far below the electron affinity EA up to a value $E'=E+h\nu$ above the EA.
Internal conversion \cite{bixon_intramolecular_1968,leger_photo_1989} quenches the electronic excitation and leads to delayed
vibrational electron detachment \cite{andersen_thermionic_2002} with an excitation-energy sensitive delay of up to $\sim$1~ms
\cite{toker_radiative_2007,goto_direct_2013,najafian_radiative_2014}.  On Co$_4^-$ clusters in an electrostatic ion beam trap (EIBT)
\cite{zajfman_electrostatic_1997,wollnik_time--flight_1990,lange_cryogenic_2010,breitenfeldt_decay_2016} we observe electron detachment
in a fixed range of delay after the laser pulse.  Operation of the trap at cryogenic temperature
\cite{lange_cryogenic_2010,menk_vibrational_2014,breitenfeldt_decay_2016} allows us to determine the range of internal cluster energies
that causes photodetatchment in the observed delay window.  Hence, from the photon energy, we know the internal energy before laser
excitation and can scan the IED by photon energy variation.

The IED is measured time-resolved (in 50~ms steps) with bin widths $<$0.1~eV.  We find near-canonical IEDs with significant deviations
for the initially hot clusters up to about 1\,s of storage at room temperature.  The average internal energies as a function of time
follow a modified Stefan--Boltzmann law.  The equilibrium internal energy and the relaxation time dependence allow conclusions on the
thermal properties of the cluster and its absorptivity for thermal radiation.

The experiments are performed with the Cryogenic Trap for Fast ion beams (CTF), an EIBT at the Max-Planck-Institut f\"ur Kernphysik in
Heidelberg \cite{lange_cryogenic_2010,breitenfeldt_decay_2016}.  Co$_4^-$ anions are produced with a metal ion sputter source (MISS)
and a laser vaporization source (LVAP).  The MISS is a cesium sputter source yielding clusters at high ro-vibrational energies
\cite{wucher_cluster_1996,wucher_internal_1998}.  In the LVAP \cite{dillinger_infrared_2015}, 10--20\,mJ pulses from the second
harmonic of a Nd:YAG laser hit a cobalt target.  Cooling the ablated plasma by a short helium pulse (14\,bar backing pressure) leads to
clustering and supersonic expansion of the helium--cluster mixture through a 2-mm nozzle yields clusters of low internal temperatures
\cite{milani_relative_1991}.

%
%
\begin{figure}[tbp]
  \centering
    \includegraphics[width=8.6cm]{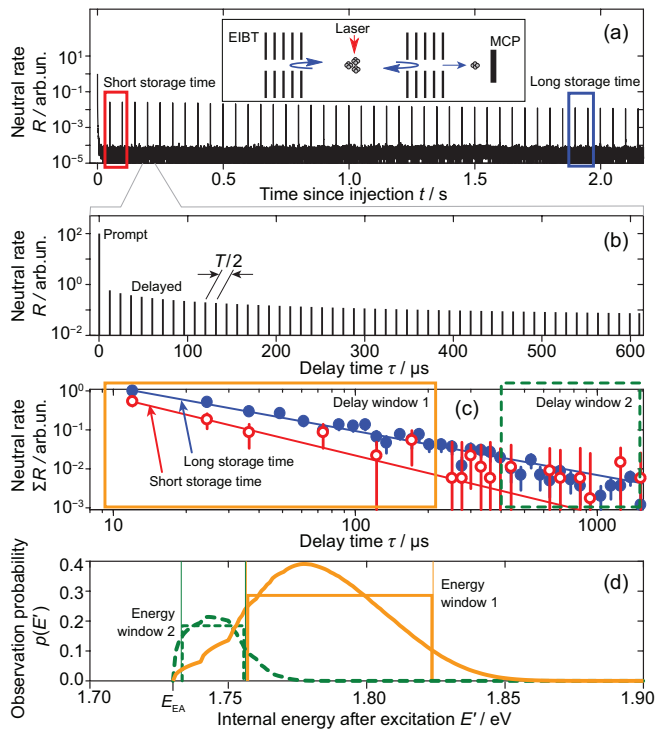}
    \caption{(a) Co$_4^-$ neutralization rate $R(t)$ (for ions produced by MISS, averaged over $\sim$10$^4$ injections) showing hot-ion
      relaxation after injection ($t=0$) and peaks induced by laser pulses with $h\nu=1.55$\,eV.  Inset: schematic of the EIBT with the
      mirror electrodes, the crossing laser, and the MCP detector. (b) Schematic laser-probing signal with delayed spikes spaced by
      half the EIBT oscillation period ($T$). (c) Summed laser signals $\Sigma R$ from laser-pulse pairs marked in (a).  Solid lines:
      approximate power-law decays (to guide the eye).  (d) Probability functions $p(E')$ for observing neutrals at the MCP (unit
      detector response assumed) as functions of the internal energy $E'$ after laser excitation [thick solid and dashed curves for the
      delay time windows marked in (c); thresholds at the detachment energy $E_{\text{EA}}$].  Rectangles: energy windows with the
      heights marking the averages of $p(E')$ in the $E'$ intervals.}
  \label{SchemaMessung}
\end{figure}
%
%

The cluster ions are accelerated to 6\,keV, mass selected by a dipole magnet and captured in the CTF.  As sketched in the inset of
Fig.\,\ref{SchemaMessung}(a), the anions oscillate with the acceleration energy over the $\sim$30~cm spacing between the electrostatic
mirrors.  Fast neutrals created by electron emission can leave the trap through the downstream mirror and are monitored by a
micro-channel plate (MCP) detector.  The count rate $R$ as a function of storage time $t$ [Fig.\,\ref{SchemaMessung}(a)] reflects the
internal cluster energies $E$ or $E'$ without or with laser excitation, respectively.  The peak at $t\sim0$, decaying within
$\sim$10~ms, is due to clusters from the hot source with internal energies $E$ above the EA.  Starting 49 ms after each injection and
repeated every 50 ms, nanosecond laser pulses from a tunable optical parametric oscillator (OPO) laser are crossed with the stored
anions in the trap center.  A pulse energy of 1--2\,mJ at $\sim$1\,cm laser beam diameter ensures dominance of single photon
absorption.

The clusters, captured as a bunch, disperse over all length and oscillation phases between the EIBT mirrors within milliseconds.  The
laser beam overlaps $\sim$50--1000, both forward and backward moving anions.  Only few of them are excited to $E'=E+h\nu$.  For
$E'>E_{{\rm EA}}$, prompt and delayed neutralization occur after a laser pulse.  While the average count rates are kept low enough to
avoid detector saturation, a signal accumulated over many injections yields the neutralization pulse structure
[Fig.\,\ref{SchemaMessung}(a)] on a background from residual-gas induced neutralization.

The fine-structure of these pulses [scheme in Fig.~\,\ref{SchemaMessung}(b)] as function of the delay $\tau=t-t_p$ after a laser pulse
at probing time $t_p$ reflects the ion oscillations (period $T=23~\mu$s).  A prompt spike comprizes neutralization within $\sim${}$T/4$
after illumination.  Then, similar to earlier experiments \cite{goto_direct_2013,najafian_radiative_2014}, spikes delayed by
$\tau=nT/2$ (integer $n$) follow from anions performing $n$ half-roundtrips before electron emission.  The spikes of $R(\tau)$
integrate delayed electron emission events that clusters undergo during $\tau \pm T/4$.  Spike amplitudes $R(\tau)$ from delayed
laser-induced emission are shown in Fig.~\,\ref{SchemaMessung}(c).  These delayed neutralization rates have the expected
\cite{andersen_thermionic_2002} approximate power-law time dependence.  For the plotted example with $h\nu=1.55$~eV, they become higher
and decrease less rapidly for longer ($\sim$1.8~s) than for short ($\sim$0.1~s) probing times.  This, at first sight counterintuitive
behaviour strongly depends on $h\nu$ as shown below.

In order to have sufficient statistics for efficient monitoring, the event yield is summed over a delay window of
$\tau=($12$\ldots$200) $\mu$s [window~1, Fig.\,\ref{SchemaMessung}(c)].  Importantly, only clusters in a limited range of internal
energy $E'$ after laser excitation show electron detachment at such delays.  With experimental parameters for delayed electron
detachment from Co$_4^-$ described below, we calculate the probability $p(E')$ to observe a neutralization event within the delay
window, finding significant likelihood only for anions with $E'$ in an $\sim$0.07~eV wide range near $\sim$1.8~eV
[Fig.~\,\ref{SchemaMessung}(d)].  We approximate $p(E')$ by a box whose height gives the average $\bar{p}$ of $p(E')$ between the sharp
limits and interpret the delayed yield to follow the number of anions excited to this $E'$ interval by the laser.  For longer delays,
$p(E')$ is downshifted and lowered.  Guided by the variation of $p(E')$ with the delay, we assign the yield in a later window~2
[$\tau = ($400$\ldots$1500) $\mu$s] to an adjacent, more narrow bin of $E'$ [Fig.~\,\ref{SchemaMessung}(d)].  Window 2 is included
since it improves the energy resolution, but as it has lower statistics, the results are dominated by window 1.

%
%
\begin{figure}[tbp]
  \centering
    \includegraphics[width=7.8cm]{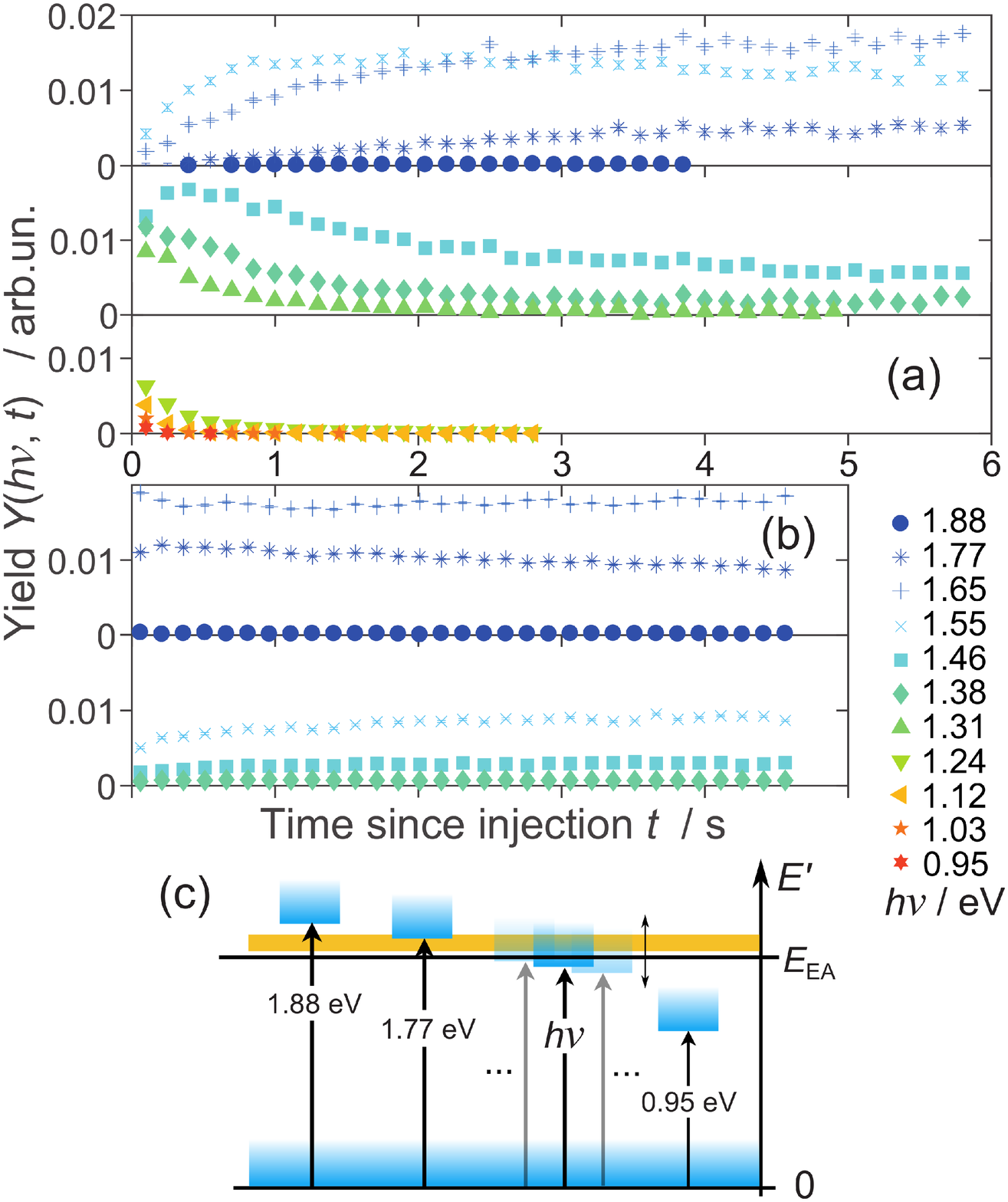}
    \caption{Normalized yield $Y(h\nu, t)$ in delay window 1 for the listed photon energies $h\nu$ and a scheme of the IED scan.  Data
      for (a) hot (MISS) and (b) cold (LVAP) ion source, averaged over three consecutive probing times. Legend: values of $h\nu$ and
      their symbols as used in (a) and (b). (c) Scheme of scanning the IED of stored ions (lower shaded band) by laser excitation at
      varying $h\nu$ (vertical arrows).  A bar just above the detachment energy $E_{\text{EA}}$ shows the sensitive range of the
      delayed photodetachment signal.}
  \label{Delayed_uber_Zeit}
\end{figure}
%
%

For window 1, Fig.~\ref{Delayed_uber_Zeit} shows the laser-induced yield $Y(h\nu,t)$, i.e., the delayed count rate normalized to the
photon number (via the laser intensity), to $\bar{p}$, and to the stored-ion number [from the background in
Fig.~\ref{SchemaMessung}(a)].  For a set of photon energies $h\nu$ (Fig.~\ref{Delayed_uber_Zeit}, legend) and the two ion sources,
$10^3$ to $8\times10^4$ injection cycles each were repeated.  As the delayed events can be associated with a specific interval of $E'$,
they reflect the IED before laser excitation, denoted by $N(E)$ with $E=E'-h\nu$.  As illustrated by Fig.~\,\ref{Delayed_uber_Zeit}(c)
the IED, shifted by one-photon absorption, is moved over the sensitive $E'$ interval by varying $h\nu$.  In fact, $h\nu$ must be low
enough for delayed yield to be observed at all: only at $h\nu\le1.77$~eV does the low-$E$ range of the shifted $N(E)$ start to overlap
with the sensitive $E'$ window.  Lower $h\nu$ probe, via the same $E'$ window, higher parts of the IED.

For the hot ion source [MISS, Fig.\,\ref{Delayed_uber_Zeit}(a)] and low $h\nu$ ($<${}$1.31$~eV), yield is found at short $t$ and later
disappears as the hot part of the IED relaxes.  Signals with $1.46\,\text{eV}< h\nu\leq1.77$~eV grow with $t$ reflecting the
increase of the IED at lower energies.  In contrast, for the cold ion source [LVAP, Fig.\,\ref{Delayed_uber_Zeit}(b)] temporal
variations are much smaller.  Yield at high $h\nu$ (small $E$) is large already at short $t$ and even slowly decreases, while at lower
$h\nu$ (higher $E$) a slow increase is observed; both indicate how initially colder ions heat up slightly during storage.

%
%
\begin{figure}[tbp]
  \centering
    \includegraphics[width=6.8cm]{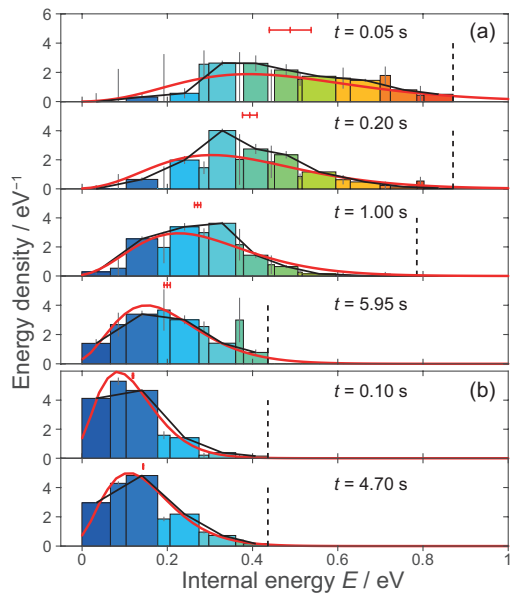}
    \caption{Normalized histograms $N(E)$ from yields $Y(h\nu,t)$ ($h\nu$ from Fig.\ \ref{Delayed_uber_Zeit}) for marked probing times,
      representing IEDs of ions produced with the (a) MISS and (b) LVAP sources.  Bin shading encodes $h\nu$.  Thin lines: statistical
      uncertainties of the bin contents.  Marks at the top of each panel: histogram averages $\bar{E}$ and their statistical
      uncertainties.  Straight segments connect the data for window~1.  Smooth curves: canonical distributions (0.02 eV bins) for
      harmonic approximation with the respective $\bar{E}$.  Dashed: upper probing energy limit.}
    \label{Energieverteilungen}
\end{figure}
%
%

The delayed electron emission from a cluster anion with internal energy above $E_{\text{EA}}$ is dominated by statistical vibrational
autodetachment \cite{andersen_thermionic_2002}.  To obtain its rate constant $k(E')$, we model this process
\cite{andersen_thermionic_2002,hansen_statistical_2013,sup}
\nocite{kafle_electron_2015,martin_fast_2013,troe_low-energy_2007,beyer_algorithm_1973,g09,walther_temperature_1999}
%
%
with vibrational-level densities of Co$_4^-$ and Co$_4$ from density functional theory.  With the level densities and $E_\mathrm{EA}$,
we derive the $k(E')$ except for a scaling factor $k_0$ which, together with $E_\mathrm{EA}$, is experimentally determined.  At
$E'\sim E_\mathrm{EA}$, where $k(E')$ is small, the competition by radiative decay of the excited Co$_4^-$ is accounted for by an
energy-independent rate constant $k_r$.  (This radiative decay also influences $N(E)$ with $E\ll E_\mathrm{EA}$ as probed by the $h\nu$
scan, but at much lower rates.)  The model parameters are determined \cite{sup} from fits of the calculated neutral-rate time
dependence to two subsets of data. The first, studied under cryogenic low-background conditions, is the initial neutral rate of hot
(MISS) Co$_4^-$ ions [signal near $t=0$ in Fig.\,\ref{SchemaMessung}(a)].  The second subset is the average of laser-induced bursts for
storage time $t>3.8$\,s.  The combined analysis yields $E_\mathrm{EA}=1.73(1)\,$eV, $k_0=240(70)\,$s$^{-1}$, and
$k_r=70(70)\,$s$^{-1}$, from which $k(E')$ and the probability functions $p(E')$ of Fig.~\,\ref{SchemaMessung}(d) are derived
\cite{sup}.  The probed internal-energy ranges are $E'=E_\mathrm{EA}+ ($0.024$\ldots$0.094)~eV for delay window 1 and
$E'=E_\mathrm{EA}+($0.003$\ldots$0.023)~eV for window 2. For probing times spaced by 50~ms, normalized histograms $N(E)$ of the IEDs
are constructed.  Here, each $h\nu$ yields a high-statistics bin from window 1 and a narrower, downshifted, but lower-statistics bin
from window 2.  Small gaps or overlaps between the bins are compensated \cite{sup} to uniquely and fully cover the $E$ scale.

Exemplary $N(E)$ are shown in Fig.\,\ref{Energieverteilungen}.  While $h\nu$ is known to $<$0.001~eV, we estimate an uncertainty of
$\pm$0.02~eV in each bin position from the uncertainty of the model for the sensitive $E'$ windows \cite{sup}.  Smoothed IED curves
using canonical level populations within the harmonic model \cite{sup}, calculated for the same average energies $\bar{E}$ as those of
the measured histograms, are superimposed.  For the hot ion source [Fig.\,\ref{Energieverteilungen}(a)], relaxation of $N(E)$ and
improving agreement with the canonical shape are seen, while for the cold ion source [Fig.\,\ref{Energieverteilungen}(b)] temporal
changes are much smaller with better fit to the canonical shape.  Lacking relevant data, we assumed the cross section $\sigma(\nu)$ for
single-photon excitation of the Co$_4^-$ anions to be constant over the investigated range of $h\nu$.  The absence of sharp resonances
appears plausible considering the broadening of electronic transitions by rotational and vibrational excitation and the expected rapid
internal conversion.  As scanning of $h\nu$ implies scanning of $E$, unaccounted structures of $\sigma(\nu)$ would cause features in
the IEDs repeating for different times at the same $E$, which is not observed.  A steady variation of $\sigma(\nu)$ along $\nu$ (and
$E$) cannot be ruled out.  With calculated canonical IEDs, we find that a possible factor-of-four \cite{sup} variation of
$\sigma(\nu)$, assumed to be linear in $h\nu$, causes a further $\pm$0.02~eV uncertainty in $\bar{E}$.  For the overall uncertainty of
$\bar{E}$ we estimate $\pm$0.03~eV.  The experimental binning is estimated (at long $t$) to upshift $\bar{E}$ by $\lesssim$0.005~eV.

\begin{figure}[tbp]
  \centering
    \includegraphics[width=7.5cm]{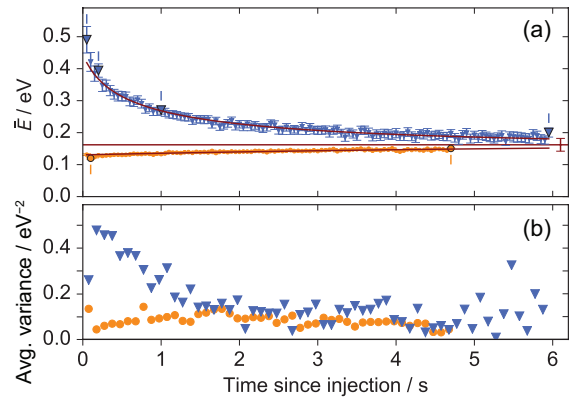}
    \caption{(a) Average internal energy $\bar{E}$ derived from the measured IEDs for ion sources MISS (triangles) and LVAP (dots) with
      statistical uncertainties.  Full curves: fit using Eq.\ (\ref{modBoltzmann}).  Horizontal line: $\bar{E}_{\mathrm{eq}}$ with the
      estimated systematic uncertainty.  Large symbols and vertical marks: data points for the IEDs of Fig.\,\ref{Energieverteilungen}.
      (b) Average variance between the IED data (bins from window 1) and the canonical energy density averaged over the same bins for
      ion sources MISS (triangles) and LVAP (dots), averaged over two consecutive probing times.}
    \label{Energie_uber_Zeit}
\end{figure}

We derive the mean internal energy $\bar{E}$ from all $N(E)$ histograms [Fig.\,\ref{Energie_uber_Zeit}(a)].  They reflect the cooling
of the hot ion ensemble and the slow heating up of the cold one.  Also, we obtain the squared deviations of the $N(E)$ bin values from
the canonical energy density in the same bins.  The average of this variance, taken over the primary (high-statistics) bins only, is
shown in Fig.\,\ref{Energie_uber_Zeit}(b).  For the hot (MISS) ensemble, initially large variances shrink and level off during the
relaxation, approaching the squared statistical uncertainty of the $N(E)$ bin contents, (0.1$\ldots$0.2)~eV$^{-2}$. For the LVAP
ensemble, the average variance is at the low value of $\sim$0.1~eV$^{-2}$ at all times.

The mean energies $\bar{E}$ for the two ion sources [Fig.\,\ref{Energie_uber_Zeit}(a)] approach a common equilibrium value.  A modified
Stefan--Boltzmann law was fitted to the data, assuming the emitted power of an ion ensemble at $\bar{E}$ as proportional to $\bar{E}^p$
($p=4$ for the Stefan--Boltzmann case) and, thus, a net power exchange of
\begin{equation}
  \frac{d\bar{E}}{dt} = B({\bar{E}_{\mathrm{eq}}}^p-\bar{E}^p)
  \label{modBoltzmann}
\end{equation}
with $\bar{E}_{\mathrm{eq}}$ the energy at equilibrium with the environment.  A combined fit of solutions of Eq.\ (\ref{modBoltzmann})
to the MISS and LVAP data yields $\bar{E}_{\mathrm{eq}}=0.162(2)_{\text{stat}}(30)_{\text{sys}}$~eV, $B=18(7)$\,eV$^{1-p}$\,s$^{-1}$,
and $p=4.2(3)$, i.e, close to the Stefan--Boltzmann value of $p=4$.

%
%
%
%

Observing that the initially hot and cold cluster ensembles both approach the same $\bar{E}$, we conclude that the internal energy
$\bar{E}_{\mathrm{eq}}$ is that of a Co$_4^-$ cluster in equilibrium with a thermal bath at the trap temperature
($T_{\mathrm{eq}}=295$\,K). The vibrational internal energy (caloric curve) of Co$_4^-$ is 0.09(1)~eV at this $T_{\mathrm{eq}}$ in the
harmonic approximation \cite{sup} allowing for a $\pm$20\% common scaling of all mode frequencies.  As the measured
$\bar{E}_{\mathrm{eq}}$ lies 0.07(3)~eV higher, it appears that harmonic vibrations alone do not adequately describe the internal
cluster energies under the conditions accessed by the present long-term observations of an isolated cluster.  In particular, it
neglects isomeric transformations \cite{ma_structures_2006,sebetci_cobalt_2008}, magnetic moments \cite{peredkov_spin_2011}, and
electronic excitation.  The latter, predicted to be $>$0.6~eV \cite{sebetci_cobalt_2008}, are likely to be negligible.

From the energy relaxation curves one can infer the absorptivity $\alpha$ of Co$_4^-$.  Equation (\ref{modBoltzmann}) with $p=4$ and
$\bar{E}\approx CT$, $C=\bar{E}_{\mathrm{eq}}/T_{\mathrm{eq}}=5.5(1.0)\times10^{-4}$\,eV/K, can be compared to the standard
Stefan--Boltzmann law $d\bar{E}/dt = \alpha A\sigma_{\mathrm{SB}}(T_{\mathrm{eq}}^4-T^4)$ with
$\sigma_{\mathrm{SB}}=3.540\times10^{-7}$~eV\,s$^{-1}$\,nm$^{-2}$\,K$^{-4}$.  Assigning to Co$_4^-$ a surface of $A\approx 0.5$~nm$^2$
results in an absorptivity of Co$_4^-$ of $\alpha \approx BC^4/A\sigma_{\mathrm{SB}} \approx (0.25\ldots2.5)\times10^{-5}$.
Alternatively, $\alpha$ can also be estimated \cite{hansen_thermal_1998} for very small particles to be of the order of their radius
($\sim$0.2\,nm for Co$_4^-$ \cite{sebetci_cobalt_2008}) divided by the wavelength ($\sim$16~$\mu$m for the peak of 295-K thermal
radiation) in agreement with our experimental order of magnitude.

In summary, we demonstrate how single-photon excitation and delay-sensitive electron detachment measurements can be combined for
laser-scanning of internal cluster energy distributions with high temporal and energetic resolution.  Further studies by this method
accessing lower equilibrium temperatures are envisaged at the CTF \cite{lange_radiative_2012} and the newly developed cryogenic storage
rings \cite{schmidt_first_2013,von_hahn_cryogenic_2016}.  With very narrow IEDs realized at these devices, laser excitation will offer
theory-independent calibration of the emission versus internal energy and, thus, application of the method to systems with less
knowledge on the relevant level densities.

We acknowledge financial support by the Max-Planck Society and the Max-Planck F\"orderstiftung. This work was supported by the German
research foundation DFG within the Transregional Collaborative Research Center SFB/TRR 88 ``Cooperative effects in homo and
heterometallic complexes'' (3MET.de).


%


\end{document}


\vspace*{-1.5cm}
{\raggedright%
  \begin{minipage}[b]{\textwidth}  {\bfseries{\small{Online
      Supplemental Material}\par}\vspace{1mm}
   \bfseries{
      {\small Long-term monitoring of the internal energy distribution of isolated cluster systems}%
    }\par}
  \end{minipage}\\[2mm]
  \begin{minipage}[b]{\textwidth} \raggedright \small
    Christian~Breitenfeldt$^{1,2}$, Klaus~Blaum$^2$, Sebastian~George$^2$,
    J\"urgen~G\"ock$^2$, Gregorio~Guzm\'an-Ram\'irez$^3$, Jonas~Karthein$^2$,
    Thomas~Kolling$^4$, Michael~Lange$^2$, Sebastian~Menk$^2$, Christian~Meyer$^2$, Jennifer~Mohrbach$^4$, Gereon~Niedner-Schatteburg$^4$, Dirk~Schwalm$^{2,5,\dagger}$, Lutz~Schweikhard$^1$, and Andreas~Wolf$^2$%
  \end{minipage}\\[1mm]%
  \begin{minipage}{\textwidth} \raggedright\itshape \footnotesize%
    $^1$Institut f\"ur Physik, Ernst-Moritz-Arndt-Universit\"at, 17487 Greifswald, Germany;
    $^2$Max-Planck-Institut f\"ur Kernphysik, 69117 Heidelberg,
    Germany;
    $^3$Departamento de Ingenier\'ias, Centro Universitario de Tonal\'a, Universidad de Guadalajara, Jal. 48525, Mexico; %
    $^4$Fachbereich Chemie, Universit\"at Kaiserslautern, 67663 Kaiserslautern, Germany; %
    $^5$Department of Particle Physics, Weizmann Institute of Science, Rehovot 76100, Israel;
    $^\dagger$deceased 14 July 2016
  \end{minipage}
  \par}
\vspace{2mm}
\hrule
\vspace{10mm}
\thispagestyle{empty}
\def\theequation{S\arabic{equation}}
\def\thefigure{S\arabic{figure}}
\def\thetable{S\arabic{table}}

\vspace{-2mm}

In this Supplemental Material, we first discuss the calculation of the delayed electron emission rate by vibrational autodetachment and
the determination of its main parameters for Co$_4^-$ from measurements.  We also discuss the reliability of the determined parameters
and present further details on the analysis procedure of the experimental internal energy distributions.
\section{Delayed vibrational autodetachment model}
\subsection{Determination  of the rate constant}

Previous studies \cite{andersen_thermionic_2002,kafle_electron_2015,martin_fast_2013,menk_vibrational_2014} have analyzed the delayed
electron emission by vibrational autodetachment as the inverse process of electron capture on a polyatomic neutral target, using a
statistical description. We apply this method to predict the decay rate constant $k$ for anionic clusters of a given internal energy
which arises either from the cluster production in the ion source or, additionally, from the one-photon excitation to $E'=E+h\nu$,
where $E$ is the anion cluster energy before excitation and $h\nu$ the photon energy.

The anionic cluster can detach its excess electron if its internal energy exceeds the adiabatic electron affinity ($E_\mathrm{EA}$) of
the neutral.  The decay constant for delayed electron emission by vibrational autodetachment can be calculated by detailed balance
\cite{andersen_thermionic_2002,hansen_statistical_2013}:
\begin{equation}
k(E')=\frac{2m}{\pi^2\hbar^3}
\int\sigma_{\text{att}}(\epsilon)\,\epsilon\,\frac{\rho_{n}(E'-E_\mathrm{EA}-\epsilon)}{\rho_{a}(E')}\,\mathrm{d}\epsilon,
\label{eq:k}
\end{equation}
with $\epsilon$ being the kinetic energy of the detached electron, $\sigma_{\text{att}}$ the cross section for attachment of an
electron to the neutral cluster, and $\rho_{n}$ and $\rho_{a}$ the vibrational energy densities of states of the neutral and anionic
system, respectively.  While electron attachment by vibrationally inelastic collisions has been calculated for better known polyatomic
systems such as SF$_6$ \cite{troe_low-energy_2007}, detailed predictions for the electron attachment cross section of Co$_4$ are not
available.  Instead, we approximate the capture cross section $\sigma_{\text{att}}$ by the Langevin expression
\cite{andersen_thermionic_2002}
\begin{equation}
\sigma_L=\pi\sqrt{\frac{2\alpha e^2}{\epsilon}},   \label{eq:sigmaL}
\end{equation}
(with $\alpha$ being the polarizability of the neutral cluster) multiplied by an energy-independent sticking coefficient $p_s$:
\begin{equation}
\sigma_{\text{att}}=p_s\sigma_L.   \label{eq:sigmaatt}
\end{equation}
The sticking probability $p_s$ takes into account that for electron attachment, energy has to be transferred from the electron to the
vibrational modes of the cluster. Thus, $p_s$ can be as low as 0.001 \cite{andersen_thermionic_2002}. As $p_{s}$ is unknown we express
the decay constant, up to an energy-independent factor, by
\begin{equation}
w(E')=\int_0^{E'-E_\mathrm{EA}}\sqrt{\epsilon}\,\frac{\rho_{n}(E'-E_\mathrm{EA}-\epsilon)}{\rho_{a}(E')}\,\mathrm{d}\epsilon.
\label{kTilde}
\end{equation}
The overall size of the decay rate constant we specify by its value $k_0$ at an anionic cluster internal energy of
$E'_0=E_\mathrm{EA}+1$\,meV. We make this choice of $E'_0$ as 1\,meV is the energy resolution of the calculations and, thus,
$E'_0 =E_\mathrm{EA}+1$\,meV the lowest energy for which $k$ is non-zero. The decay constant is then described by
\begin{equation}
k(E')=k_{0}\,\frac{w(E')}{w(E'_0)}   \label{eq:kval}
\end{equation}
where a constant factor from Eqs.\ (\ref{eq:sigmaL}) and (\ref{eq:sigmaatt}) has been taken out of the integral of Eq.\ (\ref{eq:k}).
$k_0$ then is a scaling factor with a fixed functional shape expressed by Eq.\ (\ref{kTilde}).  The experimental determination of $k_0$
and $E_\mathrm{EA}$ is described in Secs.~\ref{sec:ini} and \ref{sec:laser}.

The calculation assumes that on the time scale investigated (decay rates of $<10^6$~s$^{-1}$) the only effective process for
autodetachment is the coupling of electronic and vibrational motion (correspondingly, the cluster's internal energy in electron
attachment is via vibrational excitation only).  This leads to the concept of vibrational autodetachment
\cite{andersen_thermionic_2002}.  Electronic and rotational excitation of the Co$_4$ system are assumed not to change in this
process. Consequently, these other excitations cancel in the ratio of the densities of states for the reaction rate.  It is the large
number of vibrational levels in the neutral cluster that acts as a reservoir for the low-rate, statistical process.  Direct detachment
by electronic excitation will give rise to a prompt neutralization signal that is not considered in this experiment.  Fine-structure
excitation may contribute to the energy available for a statistical excitation of the Co$_4^-$ vibrations, but is considered here not
to directly drive electron detachment processes.  Hence, we use the vibrational energy level densities in Eq.~(\ref{eq:k}).

The required vibrational densities of states were derived with the harmonic-oscillator approach using the algorithm by Beyer and
Swinehart \cite{beyer_algorithm_1973}. The vibrational frequencies required for this step were calculated by density-functional theory
(DFT) and are listed in Table\,\ref{Parameter}.

\begin{table}[b]
  \caption{
    \label{Parameter} 
    Wave numbers (in cm$^{-1}$) of the six vibrational modes ($n$) of neutral Co$_4$ ($\nu_{n}^\text{{neut}}$)
    and the anionic Co$_4^-$ ($\nu_{n}^{\text{ani}}$).  DFT calculations were realized with the {\sc Gaussian} program \cite{g09}.  From various
    geometries and spin multiplicities tested, the geometries of
    the lowest binding energy (including the zero-point energy) and the lowest spin multiplicity were chosen (nonet for the
    neutral and decuplet for the anion).  The wave numbers of other geometries are similar in size (generally within $\sim$20\%, up to
    $\sim$60\% for $\nu_2$).}
  \vspace{2mm}
  \centering
  \begin{tabular}{c r r r r r r}
\hline
 $n$ & 1 & 2 & 3 & 4 & 5 & 6 \\
\hline
 $\nu_{n}^\text{{neut}}$& 78 & 171 & 187 & 203 & 283 & 322 \\
 $\nu_{n}^{\text{ani}}$ &85 & 138 & 210 & 217 & 296 & 304 \\
\hline
 \end{tabular}
 \end{table}

Once the energy dependent autodetachment rate can be predicted with Eq.~(\ref{eq:kval}), the neutral event rate can be calculated as a
function of time $t$ as
\begin{equation}
R(t)=C_1\int_0^{\infty}{N(E')k(E')e^{-[k(E')+k_{r}]t}\mathrm{d}E'}+C_2,
\label{RATE}
\end{equation}
where $N(E')$ is the distribution of vibrational energies in the anion and $C_{1,2}$ are constants adjusting the total scale of the
decay as well as the collisional background.  The additional rate $k_r$ takes into account radiative decays of excited levels with a
vibrational energy of $E' \approx E_{\text{EA}}$.  Typical values of $k_r$ lie between 10 and 100 s$^{-1}$
\cite{menk_vibrational_2014}.  For the case of initial internal energy of cluster anions in the ion source, $E'=E$ (distributed over
the vibrational degrees of freedom) and $t$ is the time after the ion production in the source.  For the case of an already relaxed
anionic cluster and subsequent one-photon excitation by a laser pulse, followed by redistribution of the excitation energy over the
vibrational degrees of freedom, $E'=E+h\nu$, where $E$ is the internal anionic cluster energy before the excitation and $h\nu$ the
photon energy, while $t$ is replaced by the delay time $\tau=t-t_p$ from the laser pulse at time $t_p$.

\begin{figure}[b]
  \centering
    \includegraphics[width=0.78\textwidth]{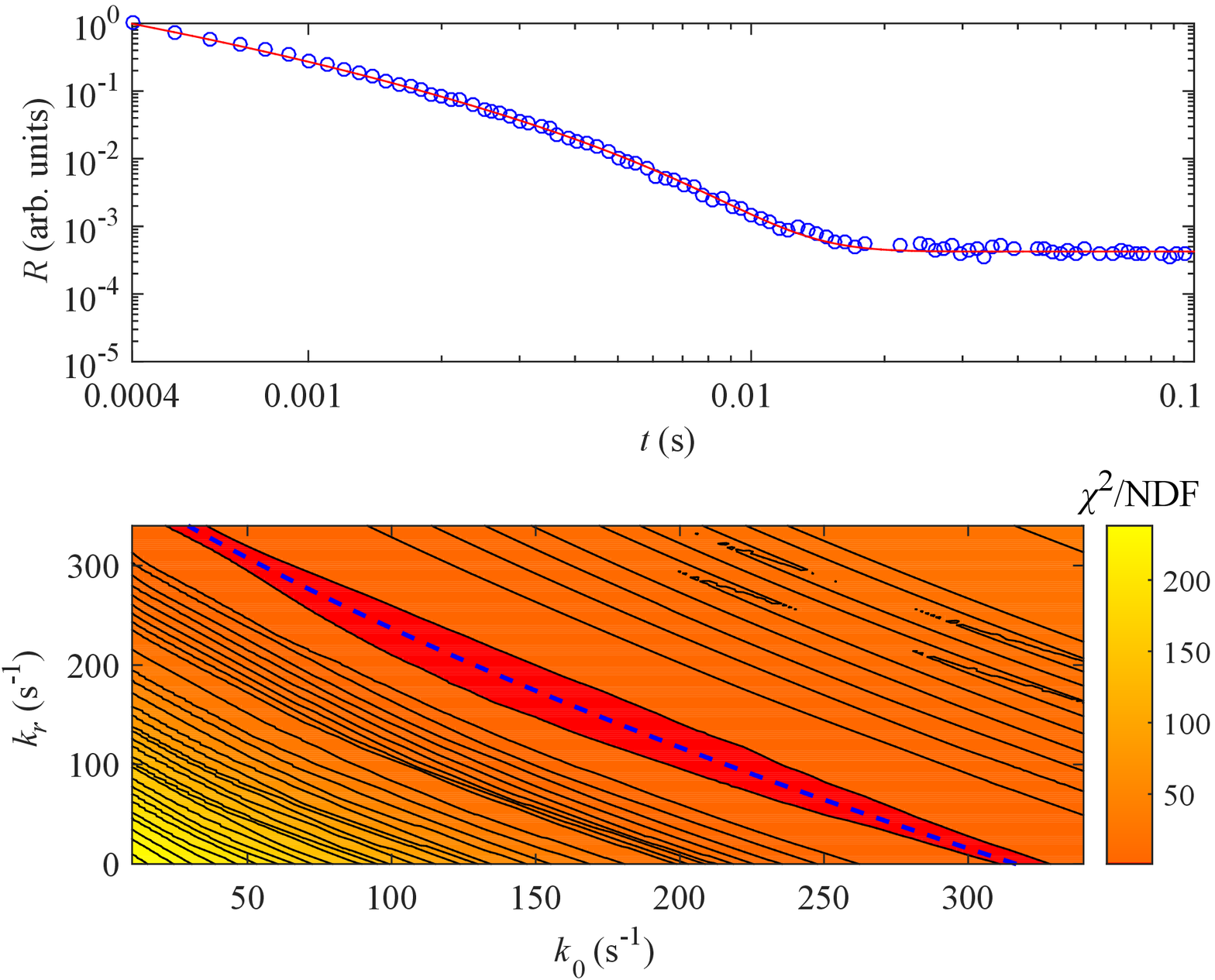}
    \caption{Top panel: Initial decay curve of Co$_4^-$ (open cicles) and fitted line (red) with $T$=1500\,K, $k_{r}=70$\,s$^{-1}$,
      $E_\mathrm{EA}$=1.73\,eV and $k_{0}=240$\,s$^{-1}$.  Bottom panel: $\chi^2$/NDF of the fit of the initial decay after
      production with the MISS for $E_\mathrm{EA}$=1.73\,eV and $T_i=$1500\,K.  The dashed blue line is the quadratic fit to the
      correlation between $k_r$ and $k_0$. (See the text for further discussion.)}
  \label{ADFit}
\end{figure}
%


%

\subsection{Analysis of the vibrational autodetachment rate: initial thermal energy}\label{sec:ini}

The initial thermal energy in the ion source results in vibrational excitation modeled by the canonic energy distribution
\begin{equation}
N(E)=C_0\,\rho_a(E)e^{-E/k_BT},  \label{eq:ni} 
\end{equation}
where $C_0$ is a normalization factor and $\rho_a(E)$ the anionic vibrational density of states.  We set $T=T_i$ with $T_i$ being the
initial (i.e., the ion source) temperature.  The time dependence of electron emission and, thus, the neutral event rate is then
calculated by Eq.\ (\ref{RATE}) with $E'=E$.

Similar to previous work on SF$_6^-$ molecules \cite{menk_vibrational_2014}, we measured the time dependence of the neutral event rate
for Co$_4^-$ with high precision (see the upper panel of Fig.\,\ref{ADFit}) under almost background free conditions achieved by
operating the EIBT at 15\,K (for this particular measurement).  This makes it possible to follow the vibrational autodetachment down to
very small rates, corresponding to excitation energies just above $E_\mathrm{EA}$ [see the upper panel of Fig.\,\ref{ADFit}].
We take care to define $t$ as the sum of the storage time and the time of flight from the ion source to the trap ($\sim 90\,\mu$s).

In fitting Eq.~(\ref{RATE}), the parameters $C_{1,2}$, $k_0$ and $k_r$ as well as $T_i$ and $E_\mathrm{EA}$ were varied.  The quality
of the fit was almost independent of $T_i$ and $E_{\text{EA}}$, tested over ranges of 1300--2500\,K and 1.65--1.78\,eV, respectively.
Regarding $k_0$ and $k_r$, the quality of fit, as represented by the reduced variance $\chi^2$/NDF, is shown in Fig.\,\ref{ADFit}
(bottom panel) (here, $T_i=1500$~K and $E_{\text{EA}}=1.73$\,eV).  Minimum $\chi^2$/NDF is obtained for combinations of $k_{0}$ and
$k_{r}$ which can be quantified by a quadratic expression (dashed blue line).  The individual values of $k_0$ and $k_r$ cannot be
extracted from these data alone.  However, their sum is approximately constant and quite precisely constrained [e.g.,
$k_0+k_r = 310(10)$~s$^{-1}$ in the expected range of $k_{r}\lesssim 100$~s$^{-1}$].  The fact that with the cryogenic EIBT the
dependence of the vibrational autoionization rate can be followed over many orders of magnitude yields the high precision of $k_0+k_r$.
Because of the broad-band excitation it can be understood that, in contrast to the laser excitation discussed below, the fit results
show little sensitivity on $E_\mathrm{EA}$.

\subsection{Analysis of the electron affinity: laser excitation}\label{sec:laser}

\begin{figure}[t]
  \centering
    \includegraphics[width=0.78\textwidth]{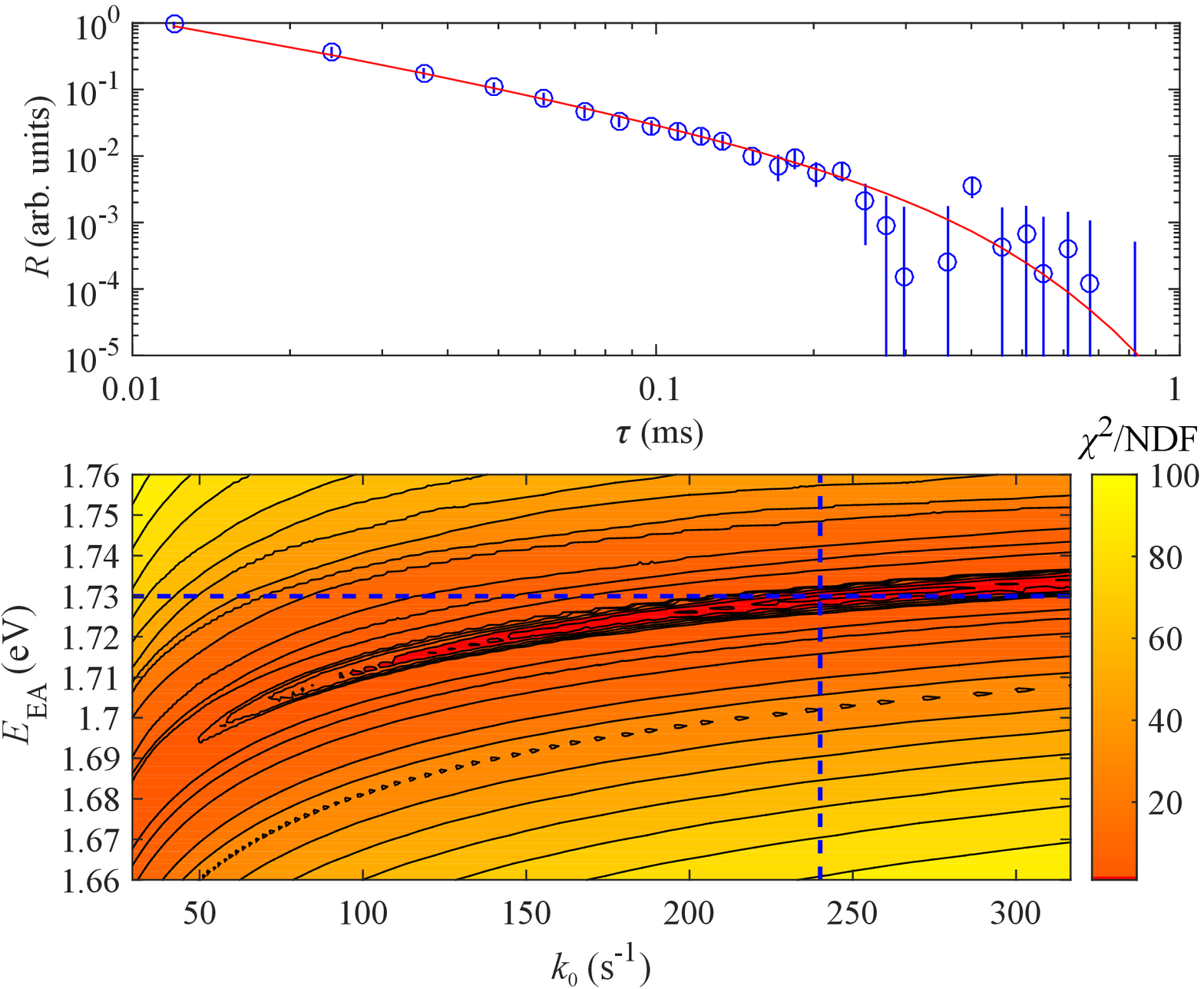}
    \caption{Top panel: Measured decay curve after excitation with photons with $h\nu=1.77\,$eV (open circles).  The ions were produced
      with the LVAP.  All data between 3.8 and 4.7\,s storage time are summed up.  The curve shows a fit with the parameters given
      below.  Bottom panel: $\chi^2$/NDF of the fitted laser induced delayed electron detachment curve for various combinations of
      the parameters $E_\mathrm{EA}$ and $k_{r}$.  The dashed horizontal and vertical lines mark $k_0=240$\,s$^{-1}$ and
      $E_\mathrm{EA}=1.73$\,eV.}
  \label{EAFit300K}
\end{figure}

The time and photon-energy dependence of the laser-induced delayed signals in Fig.~2 of the main paper allow to delimit the value of
$E_{\text{EA}}$.  The time dependence of the neutral rate follows Eq.~(\ref{RATE}) with $E'=E+h\nu$.  At long storage times ($t$), when
the yields $Y$ in delay window 1 become similar for both ion sources [main paper, Fig.~2(a) and (b)], the laser-induced signal as a
function of $h\nu$ first appears when going down from $h\nu=1.88$~eV to 1.77~eV.  We conclude that, after one-photon absorption at
1.88~eV, $E'$ is too high (and the decay too fast, even for the smallest initial energy $E$) to yield a signal in delay window 1.
However, for $h\nu=1.77$~eV some anions with initial energies $E\sim0$ overlap after laser excitation with the energy band which is
sensitive to delayed autodetachment [main paper, Fig.~2(d)].  The decay signal is expected, on the one hand, to depend on the position
of the lower edge of the shifted energy distribution (near 1.77~eV) with respect to $E_{\text{EA}}$.  On the other hand, only the
lowest part of the shifted energy distribution lies in the sensitive range (corresponding to the observed decay times of
$\ge${}12~$\mu$s) and, hence, the decay behaviour at these times will depend only little on the thermal energy of the anions before
excitation.  We hence analyze the time dependence for $h\nu=1.77$~eV and use the data with the highest yield at this $h\nu$, taken for
the cold (LVAP) ion source (see Fig.~2 of the main paper).  To ensure sufficient statistics and to minimize storage-time dependent
fluctuations, we average the laser-induced decay time dependences over all LVAP data for the nearly stable conditions at later storage
times ($>$3.8\,s).

To the data (see Fig.\,\ref{EAFit300K}, upper panel) we fit the time dependence of Eq.~(\ref{RATE}) by adjusting $C_1$ and $C_2$ while
varying $E_\mathrm{EA}$ from 1.66\,eV to 1.77\,eV and $k_0$ from 30\,s$^{-1}$ to 315\,s$^{-1}$ (see bottom panel of
Fig.\,\ref{EAFit300K}).  The range of small $\chi^2$/NDF is rather well defined regarding $E_{\text{EA}}$.  Especially, using the
result from the cryogenic measurement of $k_0+k_r = 310(10)$~s$^{-1}$ and considering the expected range of
$k_{r}\lesssim 100$~s$^{-1}$ (i.e., $k_{0}\gtrsim 210$~s$^{-1}$), we find $E_\mathrm{EA}=1.73(1)$~eV.  Note that this calibration
yields the result that $E_\mathrm{EA}$ lies below the well known photon energy of $h\nu=1.77$~eV by 0.04(1)~eV.  A shallow minimum of
$\chi^2$/NDF as a function of the parameter $k_0$ shows that $k_0=240(70)$~s$^{-1}$, which implies $k_r=70(70)$~s$^{-1}$.  The
uncertainty in $k_0$ has little influence on the derived internal energy distribution (estimated below), since the latter is measured
through a scan of the laser frequencies.

The DFT calculations resulted in a polarizability of $\alpha=29$\,\AA.  Using this value to calculate the sticking probability results
in $p_{s}=3(1)\%$, which is a reasonable value \cite{andersen_thermionic_2002}.


\section{Experimental internal energy distributions}

\subsection{Detection probability function and energy bins}
\label{sec:prob} 

\begin{figure}[t]
  \centering
    \includegraphics[width=0.78\textwidth]{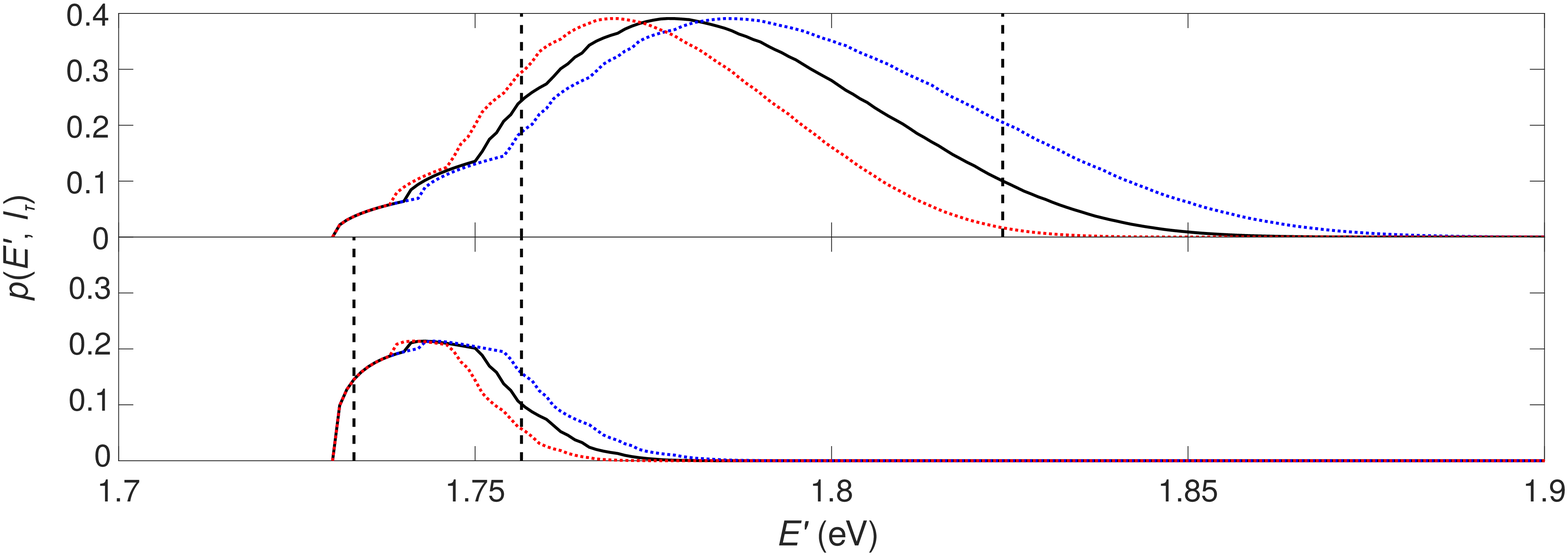}
    \caption{Probability functions $p(E',I_\tau)$ for $I_{\tau,1}=[12\,\mu$s$,200\,\mu$s] (upper panel) and
      $I_{\tau,2}=[400\,\mu$s$,1500\,\mu$s] (lower panel). The black curves are identical to Fig.~1(d) of the main paper.  The bin
      limits (dashed vertical lines) belong to these curves.  For the red (blue) curves the vibrational frequencies were reduced
      (increased) by 20\%.}
  \label{Dos}
\end{figure}

For cluster anions excited to energy $E'$ we give the probability of leading to a neutralization event at the detector in the time
interval $I_\tau=[\tau_1,\tau_2]$ after a laser pulse as
\begin{eqnarray}
p(E',I_\tau)&=&\frac{1}{2}\int_{\tau_1}^{\tau_2}k(E')e^{-[k(E')+k_r]\tau}\mathrm{d}\tau \nonumber \\
&=&\frac{k(E')}{2[k(E')+k_r]}\left( e^{-[k(E')+k_{r}]\tau_1} - e^{-[k(E')+k_{r}]\tau_2} \right).
\end{eqnarray}
As functions of $E'$ these probability functions are shown in Fig.\ \ref{Dos} [see also Fig.~1(d) of the main paper].  The energy bin
assigned if an event is detected within a given time window is defined as follows.  First, limits are found that contain 95\% of the
area for each the $p(E')$ curves that belong to delay window 1 or 2, respectively.  Window 1 leads to the upper energy ($E'$) interval
and window 2 to the lower one (Fig.~\ref{Dos}).  The upper limit of the upper $E'$ interval and lower limit of the lower $E'$
interval (``outer limits'') are retained.  At the inner limits, the energy intervals overlap; hence, the upper limit of the lower
interval is shifted down and the lower limit of the upper interval shifted up until they are only 1~meV apart and exclude equal
relative amounts of the $p(E')$ areas.  This leads to the limits listed in Table~\ref{windows}.  The relative amounts of area excluded
by the inner limits are 13\% for each of the two intervals.

\begin{table}[b]
  \caption{ 
    Windows in the delay time, $I_\tau$, the energy after excitation, $I_{E'}$ for $E_{\text{EA}}=1.73$ eV, and the energy
    windows relative to $E_{\text{EA}}$, $I_{E'-E_{\text{EA}}}$.  The limits of $I_\tau$ are given in $\mu$s 
      and those of $I_{E'}$ and $I_{E'-E_{\text{EA}}}$ in eV.  
      The parameter uncertainties yield (see Fig.~\ref{Dos}) a $\pm$20\% uncertainty for the limits of $I_{E'-E_{\text{EA}}}$
      (third column).
      \label{windows}  }
    \vspace{2mm}
    \centering
  \begin{tabular}{c c c c}
    \hline
    Window no. & $I_\tau$ & $I_{E'}$ & $I_{E'-E_{\text{EA}}}$ \\
    \hline
    1 & $[12,200]$ & $[1.757,1.824]$  &  $[0.024,0.094]$  \\
    2 & $[400,1500]$ & $[1.733,1.756]$& $[0.003,0.023]$   \\
    \hline
  \end{tabular}
\end{table}

In this derivation, the energies are essentially defined relative to $E_{\text{EA}}$ as given by the last column of Table~\ref{windows}.
The shifts of the limits occurring when the vibrational frequencies of the model are reduced or increased by 20\% are illustrated by
Fig.\ \ref{Dos}.  Approximately, the bin limits of $I_{E'-E_{\text{EA}}}$ shift by about $\pm$20\%.

The intervals of internal anionic cluster energies $E$ are obtained from the limits of $I_{E'-E_{\text{EA}}}$ by subtracting the
quantity $h\nu-E_{\text{EA}}$.  Hence, the energy accuracy for the bins of the internal energy distribution is given by the accuracy in
$E_{\text{EA}}$, amounting to $\pm$0.01~eV.  Through a $\pm$20\% inaccuracy in the vibrational frequencies the average position of the
bin related to window 1 may in addition shift by about $\pm$0.01~eV.  From this, we state an estimated overall uncertainty of about
$\pm$0.02~eV for the internal energy bins.  (Clearly, values $E<0$, which may occur for $h\nu=1.77$~eV, are eliminated from the
analysis by letting the corresponding bin start at $E=0$.)  

We have also investigated the sensitivity of the complete analysis procedure leading to the final average internal energy
$\bar{E}_{\text{eq}}$ [Eq.\ (1) of the main paper] on the precise choice of the parameter $k_0$.  By varying $k_0=240(70)$~s$^{-1}$
within its uncertainty range (between 160~s$^{-1}$ and 300~s$^{-1}$), we find a variation of $\bar{E}_{\text{eq}}$ between 0.167~eV and
0.158~eV, respectively.  ($\bar{E}_{\text{eq}}=0.162$~eV for $k_0=240$~s$^{-1}$.)  From this, estimate that the uncertainty at which
$k_0$ can be experimentally determined leads to a $\pm$3\% uncertainty of the internal energy scale ($\pm$0.005 eV on
$\bar{E}_{\text{eq}}$).  

Taken together both influences discussed in this Section, we state the systematical uncertainty of the internal energy through the
model for the sensitive $E'$ windows as $\pm$0.02~eV.  As described below, possible unaccounted variations of the photon absorption
cross section lead to additional independent systematic uncertainty of about the same size.  From this, we quote the systematic uncertainty of the
final average internal energy $\bar{E}_{\text{eq}}$ as $\pm$0.03 eV.  All uncertainties given are to be understood as single standard
deviations.

\subsection{Shape of the energy distribution function and photon absorption cross section}

Experimental parameters such as the detection efficiency do not affect the distribution function as they can be assumed to be
independent of the photon energy and constant during the measurement; their absolute amount is eliminated by normalizing the binned
internal energy distributions in a final step.  Moreover, normalization to variations in the stored ion is assured within an estimated
uncertainty of 20\%.

The largest uncertainty in the shape of the internal energy distributions is due to the photon absorption cross section $\sigma(\nu)$
of Co$_4^-$, which is assumed to be independent on the photon energy in the range of $\sim$1.8 eV to $\sim$1.1 eV.  The functional
dependence $\sigma(\nu)$ influences both the derived energy distributions (Fig.\,3 of the main paper) and the derived mean energy (main
paper, Fig.\,5).

No experimental absorption cross sections $\sigma(\nu)$ are available for this system.  For comparable metal clusters, an energy
dependence was observed.  Thus, in case of V$_{13}^+$ the cross section increases with photon energy \cite{walther_temperature_1999},
the data being well reproduced by Mie theory.  Applying Mie theory to Co$_4^-$, the cross section is predicted to monotonously increase
by a factor of four over the photon energy range investigated.  In contrast, for Al$_4^-$ a cross section decreasing with photon energy
was observed \cite{kafle_electron_2015}.  Sharp resonances are unlikely to occur considering the vibrational configuration space and
the high density of vibrationally excited levels.
%

As the energy bins in $E$ are the constant intervals $I_{E'}$ from Table~\ref{windows}, shifted down by $h\nu$, any given energy $E$
corresponds to a certain value of $h\nu$.  Therefore, variations of $\sigma(\nu)$ should produce the same $E$-dependent modification factor
for each of the distributions in Fig.\ 3 of the main paper.  As no clear repeating structures are observed in the various distributions
and lacking better options, we choose to assume a constant photon excitation rate.  However, smooth factor-of-four variations of
$\sigma(\nu)$ could occur as mentioned above.  

To model possible effects of a smooth variation of $\sigma(\nu)$, we calculated its possible effect on a canonical energy distribution
near the measured equilibrium internal energy $\bar{E}_{\rm eq}$.  We calculate these distributions from Eq.\ (\ref{eq:ni}) with the
vibrational level density $\rho_a(E)$ from the harmonic parameters of Table~\ref{Parameter} and apply the experimental binning.
Calculating the average of $E$ with this binned distribution we find, at $T=295$~K, $\bar{E}=0.0942$~eV.  The unbinned average (i.e.\
with the 1~cm$^{-1}$ binning of the calculated $\rho_a(E)$, negligibly small compared to the experimental bins) we find as
$\bar{E}=0.0896$~eV.  Hence, the experimental binning alone shifts the measured average energy up compared to the unbinned case by
0.0046 eV.

We then define a correction factor of 1 at $h\nu=E_{\text{EA}}=1.73$ eV ($E=0$) and of either 1/4 or 4 at $h\nu=0.90$~eV ($E=0.83$~eV),
varying linearly with $E$ between these limits, to simulate the possible variations of $\sigma(\nu)$ according to the previous
discussion.  We multiply the bin contents with the correction factors for both cases, taken at the bin centers, and calculate the
average $E$ of the corrected distributions.  At $T=295$~K, $\bar{E}$ changes by $-0.0036$~eV for a reduction and $+0.0134$~eV for an
increase of the correction factor along $E$ [hence, a change of $\sigma(\nu)$ as $\nu$ decreases].  In the region of
$\bar{E}_{\rm eq}=0.162$~eV (canonical $T=436$~K) the binning effect is $+$0.0043~eV and the shifts by factor-of-four sensitivity
variations are $-0.0083$~eV and $+0.0180$~eV for a reduction and increase along $E$, respectively.  We approximate our uncertainty
estimate through a factor-of-four variation of $\sigma(\nu)$ by a symmetric range of $\pm$0.02~eV.

\subsection{Binning the energy distribution}

\begin{figure}[t]
  \centering
    \includegraphics[width=0.5\textwidth]{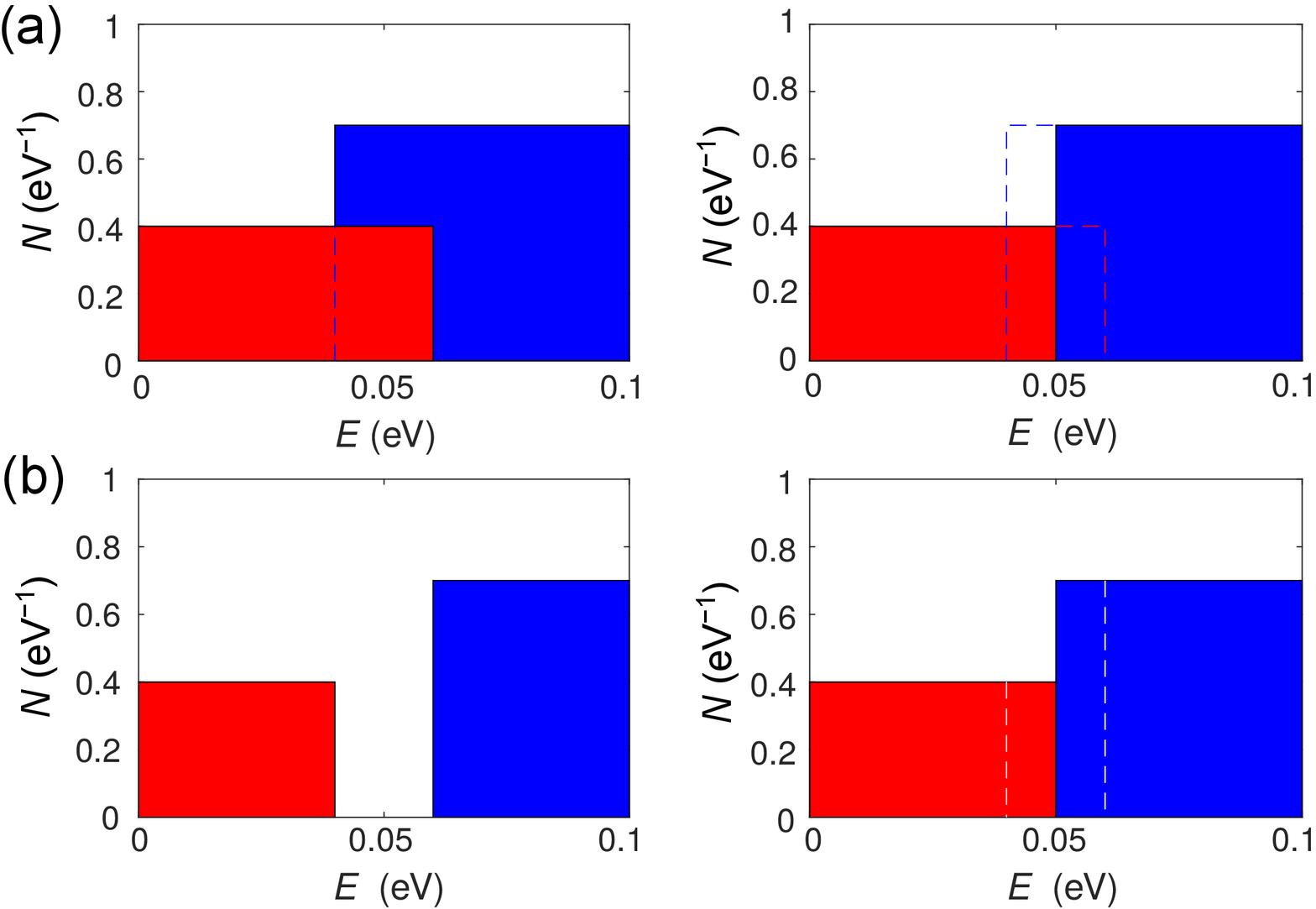}
    \caption{Redefinition of the energy bins to avoid overlaps (a) and gaps (b) in the energy distribution. The left and right diagrams
      show the bin boundaries before and after the adjustment, respectively. See text for details.}
  \label{reconstructing}
\end{figure}

The photon energies used were non-equidistant.  Thus, considering the bins of the internal energy distribution function, some ranges of
$E$ are not probed at all, while others are probed by two different wavelengths.  For reconstructing the energy distribution
continuously and uniquely, the following adjustments were made: When the upper edge of the previous bin in $E$ lies above the lower
edge of the next one, as in Fig.\,\ref{reconstructing}(a), a new boundary where the two bins are touching each other is placed half way
within the overlap region, reducing the width of each bin by the same amount.  Conversely, when the upper edge of the previous bin lies
below the lower edge of the next one, Fig.\,\ref{reconstructing}(b), a new boundary where the two bins are touching each other is
placed half way within the gap, increasing the width of each bin by the same amount. In both cases, the value of the distribution
function within the bin is left unchanged.  In a final step the energy distributions where normalized by applying a factor to the energy distribution such that
\begin{equation}
  \sum_{\tilde{I}_E} N(\tilde{I}_E)\Delta E(\tilde{I}_E) = 1
\end{equation}
with the adjusted bins $\tilde{I}_E$.

%
